# Two-Gap Superconductivity and Decisive Role of Rare-Earth *d* Electrons in Infinite-Layer Nickelates

Zhenglu Li[1,2,3] and Steven G. Louie[1,2,*]

[1]*Department of Physics, University of California at Berkeley, Berkeley, CA, USA.*

[2]*Materials Sciences Division, Lawrence Berkeley National Laboratory, Berkeley, CA, USA.*

[3]*Mork Family Department of Chemical Engineering and Materials Science, University of Southern California, Los Angeles, CA, USA.*

*Email: sglouie@berkeley.edu

**Abstract**

We present a theoretical prediction of a phonon-mediated two-gap superconductivity in infinite-layer nickelates $Nd_{1-x}Sr_xNiO_2$ by performing *ab initio GW* and *GW* perturbation theory calculations. Electron *GW* self-energy effects significantly alter the characters of the two-band Fermi surface and enhance the electron-phonon coupling, compared with results based on density functional theory. Solutions of the fully **k**-dependent anisotropic Eliashberg equations yield two dominant *s*-wave superconducting gaps – a large gap on a band of rare-earth Nd *d* and interstitial orbital characters and a small gap on a band of transition-metal Ni *d* character. Increasing hole doping induces a non-rigid-band response in the electronic structure, leading to a rapid drop of the superconducting $T_c$ in the overdoped regime in agreement with experiments.

Still holding the record for the highest superconducting $T_c$ under ambient conditions, cuprates remain among the most mysterious materials in condensed matter physics, despite extensive studies in the last four decades [1]. In the pursuit of new superconductors as cuprate analogs, the recent success in discovering superconducting infinite-layer nickelates with rather high $T_c$ raises the hope of gaining more insights into unconventional superconductivity [2-5]. Infinite-layer nickelates and cuprates share a similar transition-metal-oxygen planar structure motif, and Ni$^+$ and Cu$^{2+}$ are in the same $3d^9$ atomic valence configuration in the undoped parent compounds. Upon hole doping of $x = 0.1 - 0.3$ in Nd$_{1-x}$Sr$_x$NiO$_2$, superconductivity is observed [1-5] and reaches a highest reported $T_c$ of 23 K at $x = 0.15$ [5]. To date, several nickelate superconductors have been discovered [2-5,6-9], including Nd$_{1-x}$Sr$_x$NiO$_2$, Pr$_{1-x}$Sr$_x$NiO$_2$, La$_{1-x}$Sr$_x$NiO$_2$, La$_{1-x}$Ca$_x$NiO$_2$, and Nd$_6$Ni$_5$O$_{12}$. The similarity in the crystal structure and the presumed electronic structure of these materials to the cuprates [10,11] lead to a prevailing perspective of superconductivity in the infinite-layer nickelates as being unconventional and mainly originated from the Ni $d_{x^2-y^2}$-band [2,12-15].

While the nature of superconductivity in nickelates remains under debate, the conventional mechanism of pairing due to electron-phonon ($e$-ph) coupling was not supported from early on [12,13,15]. Previous DFT calculations showed that the $e$-ph coupling constant λ in NdNiO$_2$ is only ~0.2, accounting for a phonon-mediated $T_c <$ 1 K [16]. Existing theoretical model studies [17-22] on exploring unconventional mechanisms (typically with assumed interaction forms, downfolded subspaces, and parametrized coupling strengths) are mostly focused on the Ni $d$ band. Meanwhile, scanning tunneling spectroscopy (STS) experiment [23] on nickelate thin films interestingly observed two types of superconducting gaps (as well as their mixture) depending on the tip position in the measurements: one is a V-shape d$I$/d$V$ profile which has been interpreted as an unconventional $d$-wave gap, and the other is a U-shape profile which is a typical signature of an $s$-wave gap. Moreover, recent superfluid density experiments [24,25] revealed both nodal and fully-gapped behaviors in infinite-layer nickelates. These intriguing results do not provide a straightforward and self-consistent picture for superconductivity in the infinite-layer nickelates.

It is important to note that Kohn-Sham orbitals of DFT are not constructed to, and often may not, describe well the true quasiparticle excitation energies (e.g., the band structure) [26-29] as well as the $e$-ph coupling strength [30-33]. Thus, conclusions based on DFT calculations may be questionable, especially for materials with non-negligible electron correlations. On the other hand, the *ab initio GW* approach [26,34,35] has achieved much success in describing, from first principles, the quasiparticle properties of many materials [26,28,29] including the $e$-ph coupling [30,31]. Moreover, the fully **k**-dependent anisotropic Eliashberg theory [36-39] has become a standard computational method for solving for superconductivity in real materials [33], e.g., MgB$_2$ [40] and hydrides [41]. However, with a few exceptions, virtually all existing *ab initio* anisotropic Eliashberg theory calculations are performed in conjunction with DFT approaches [33,38-41]. The *GW* band structures and *e*-ph coupling, while being more accurate, have not been fully exploited in use with anisotropic Eliashberg theory.

Here, by combining the *GW* band structures and *GW* *e*-ph interactions with the anisotropic Eliashberg theory, our *ab initio* calculations reveal a phonon-mediated *s*-wave two-gap superconductivity in $Nd_{0.8}Sr_{0.2}NiO_2$ with a theoretically predicted $T_c$ of ~ 22 K (experimental $T_c$ = 19.3 K at $x$ = 0.2 [5]). We hence arrive at an opposite conclusion from that drawn from standard DFT results on the role of phonons in infinite-layer nickelate superconductors. We find that many-electron (self-energy) effects rearrange the bands near the Fermi energy ($E_F$) and drastically reshape the Fermi surface (FS) and its orbital characters compared with DFT – by introducing a substantial amount of Nd $d_{z^2}$ and interstitial orbital (IO) components [16,21,42] at $E_F$. The Nd-IO states host strong *e*-ph coupling strength. Our computed results show that the two-band FS exhibits two distinct *s*-wave superconducting gaps, with characteristic values of 4.7 meV on the Nd-IO band and 2.1 meV on the Ni band. The predicted two-gap superconductivity provides a strong explanation of the intriguing observations of multiple forms of gap behavior in the tunneling [23] and superfluid density [24,25] experiments by considering inhomogeneity in sample quality. Moreover, upon increasing hole doping, the Nd-IO band rapidly moves away from $E_F$, explaining the decrease of $T_c$ towards the overdoped regime as observed experimentally [3-5].

Fig. 1(a) and 1(b) show the DFT (with generalized gradient approximation (GGA)) and *GW* band structures of $Nd_{0.8}Sr_{0.2}NiO_2$, respectively, interpolated and analyzed with maximally localized Wannier functions (MLWFs) [43]. Two bands cross $E_F$, where one band is of Ni $d_{x^2-y^2}$ character (Fig. 1(f)), and the other band is mostly of Nd $d_{z^2}$ character (Fig. 1(g)) and of IO character (Fig. 1(h)) centering around a hollow site coplanar with the Nd atoms, with some hybridizations with other characters (e.g., the Ni $d_{z^2}$ and Nd $d_{z^2}$ orbitals hybridize near the Γ point). We denote these two bands by their dominant characters as the Ni band and the Nd-IO band. Comparing with DFT-GGA results, the *GW* self-energy effects bring significantly more Nd-IO characters to the FS, by shifting the Nd-IO band (Ni band) downward (upward) relative to $E_F$. The Ni $d_{x^2-y^2}$ orbital is more spatially localized than the Nd $d_{z^2}$ and the IO orbitals (Fig. 1(f-h)), therefore the two bands experience different electron self-energy effects originating from the Coulomb interaction. Refined DFT exchange-correlation functionals such as meta-GGA or hybrid functional [44,45], as well as *GW* plus dynamical mean-field theory (DMFT) [46], also give band renormalization effects following a similar trend as the *GW* results. Our results further show that the self-energy effects (more Nd-IO characters for states at $E_F$) are robust within typical *GW* numerical uncertainties across the hole-doping regime (Supplemental Material [47]). We note that the renormalized band structures vary with different rare-earth elements in infinite-layer nickelates [46,94]. Fig. 1(i) shows the projected density of states (DOS) of $Nd_{0.8}Sr_{0.2}NiO_2$ from DFT-GGA and *GW*. Unlike the case of DFT-GGA where the Ni $d_{x^2-y^2}$ states dominate the DOS at $E_F$, the Nd-IO components are comparable with Ni $d$ states in the $GW$ results, which as discussed below is critical for having two gaps and a significant high $T_c$ in phonon-mediated superconductivity.

The microscopic theory of superconductivity based on $e$-ph coupling makes use of the quasiparticle band energy $\varepsilon_{n\mathbf{k}}$ and the $e$-ph matrix element $g_{mn\nu}(\mathbf{k}, \mathbf{q})$ – the scattering amplitude between the states $|n\mathbf{k}\rangle$ and $|m\mathbf{k} + \mathbf{q}\rangle$ induced by the electron potential change from the phonon mode $|\mathbf{q}\nu\rangle$ – among other ingredients. Here, $m$ and $n$ are electron band indices, $\mathbf{k}$ and $\mathbf{q}$ are wavevectors, and $\nu$ is the phonon branch index. These ingredients can be computed from first principles [33]. The prevalent *ab initio* approaches have been based on band structures and $e$-ph matrix elements from DFT methods, which however, do not describe true quasiparticle properties and can fail qualitatively in more correlated materials [28,29]. The $GW$ method for quasiparticle band energies [26] and the recently developed $GW$ perturbation theory ($GW$PT) [30,31] for $e$-ph matrix elements properly account for the self-energy effects in materials. They provide more accurate ingredients for the first-principles computation of $e$-ph coupling and phonon-mediated superconductivity. In this work, we denote the DFT results as those computed with $\varepsilon_{n\mathbf{k}}^{\text{DFT}}$ and $g_{mn\nu}^{\text{DFT}}(\mathbf{k}, \mathbf{q})$, and the $GW$ results as those computed with $\varepsilon_{n\mathbf{k}}^{GW}$ and $g_{mn\nu}^{GW}(\mathbf{k}, \mathbf{q})$, unless otherwise stated.

Fig. 2(a) and 2(b) show the state-resolved (band- and wavevector-resolved) $e$-ph coupling strength [37-39] $\lambda_{n\mathbf{k}} = \sum_{m\nu\mathbf{q}} |g_{mn\nu}(\mathbf{k}, \mathbf{q})|^2 \delta(\varepsilon_{m\mathbf{k}+\mathbf{q}} - E_\text{F}) \times \frac{2}{\hbar\omega_{\mathbf{q}\nu}}$ for states near $E_\text{F}$, where $\omega_{\mathbf{q}\nu}$ is the phonon frequency. $\lambda_{n\mathbf{k}}$ gives the $e$-ph coupling strength for the state $|n\mathbf{k}\rangle$ coupling with all electronic states on the FS induced by all phonon modes. The Nd-IO states have much stronger $e$-ph coupling strength than the Ni states, in both DFT and $GW$ calculations. Electron correlation (through the $GW$ self-energy) notably gives rise to two important effects: 1) it brings more states of Nd-IO characters (with stronger $e$-ph coupling) to the FS (Fig. 1); and 2) it significantly enhances the magnitude of the $e$-ph matrix elements. Fig. 2(c) shows the Eliashberg function $\alpha^2 F(\omega)$ which describes the phonon-frequency dependent $e$-ph coupling strength. The $GW$ $e$-ph coupling is remarkably higher than that of DFT basically at all frequencies. Consequently, the Fermi-surface averaged $e$-ph coupling constant [33,37] $\lambda = \frac{1}{N_\text{F}} \sum_{n\mathbf{k}} \lambda_{n\mathbf{k}} \delta(\varepsilon_{n\mathbf{k}} - E_\text{F})$ (where $N_\text{F}$ is the total DOS at $E_\text{F}$) is 0.71 at the $GW$ level but only 0.13 at the DFT level, *showing an enhancement by a factor of 5.5*. Experimentally, a temperature-dependent Hall effect was observed in Nd$_{1-x}$Sr$_x$NiO$_2$ [3,4], indicating a two-carrier (of very different characters) transport scenario, which is consistent with the theoretical results of a hole-like and an electron-like band at $E_\text{F}$. Due to the stronger $e$-ph coupling in the Nd-IO band, combining with any residual scattering mechanisms (such as defects), the mobility of the Nd-IO band is suppressed more significantly than that of the Ni band at elevated temperatures with increased phonon population. Moreover, an earlier DMFT study [95] revealed that the electron-electron self-energy for the Ni $d_{x^2-y^2}$ states is strongly temperature dependent. These self-energy effects from both the $e$-ph coupling and electron-electron interaction, together with defect scattering, provide a conceptual understanding of the measured temperature-dependent Hall effect. Future studies are however needed for a more detailed understanding of the transport behaviors of infinite-layer nickelates.

In this study, we construct the fully anisotropic Eliashberg equations [36-39] using the $GW$ quasiparticle band structure and the $GW$PT $e$-ph matrix elements to solve for the superconducting properties of $Nd_{1-x}Sr_xNiO_2$. The **k**- and **q**-dependence is densely sampled via Wannier interpolation techniques [43,96]. We have also considered possible short-range antiferromagnetic fluctuations of local Ni moments on the electronic structure and find that they have little effects on the central Nd-IO band [47].

Fig. 3(a) shows the distribution of the superconducting gap [33,36-40] $\Delta_{n\mathbf{k}}$ on the FS of $Nd_{0.8}Sr_{0.2}NiO_2$ at temperature $T = 5$ K, obtained by solving the anisotropic Eliashberg theory at the full $GW$ level. The effective Coulomb potential [33,37-40] $\mu^*$ is set to 0.05, and the dependence in $\mu^*$ for superconducting properties such as $T_c$ is very weak [47]. The gap values are highly band- and **k**-dependent and vary dramatically on the FS, leading to a bimodal distribution of gaps (Fig. 3(b)) – i.e., $Nd_{0.8}Sr_{0.2}NiO_2$ hosts a fundamentally two-gap superconductivity. Fig. 3(b) shows the density distribution of $\Delta_{n\mathbf{k}}$ with orbital projections. The Ni band has small gaps of 0.5 – 2.5 meV with an average of 1.7 meV, whereas the Nd-IO band has large gaps of 3 – 5 meV with an average of 4.1 meV.

To understand the origin of the two-gap nature of the superconductivity, we plot in Fig. 3(c) the density distribution of the state-pair-resolved $e$-ph coupling strength [38,40] $\lambda_{n\mathbf{k},m\mathbf{k}+\mathbf{q}} = N_F \sum_\nu \frac{2}{\hbar\omega_{\mathbf{q}\nu}} |g_{mn\nu}(\mathbf{k},\mathbf{q})|^2$, which represents the $e$-ph coupling strength between the pair of states $|n\mathbf{k}\rangle$ and $|m\mathbf{k}+\mathbf{q}\rangle$ induced by all phonon branches. The intra-band $e$-ph coupling distributions are quite distinct for the Ni band and the Nd-IO band. The Ni ↔ Ni coupling is narrowly distributed at low coupling values (with an average $\lambda^{\text{avg.}}_{\text{Ni} \leftrightarrow \text{Ni}} = 0.22$) whereas the Nd-IO ↔ Nd-IO coupling is broadly distributed up to a high coupling value (with $\lambda^{\text{avg.}}_{\text{Nd-IO} \leftrightarrow \text{Nd-IO}} = 1.88$ and some individual values up to ~4.0). The Nd-IO ↔ Ni inter-band $e$-ph coupling shows an asymmetric distribution within a relatively moderate coupling range (with $\lambda^{\text{avg.}}_{\text{Nd-IO} \leftrightarrow \text{Ni}} = 0.58$). These distinctive distributions of $\lambda_{n\mathbf{k},m\mathbf{k}+\mathbf{q}}$ result in a well separated two-peak structure in the state-resolved $\lambda_{n\mathbf{k}}$ density distribution (with $\lambda^{\text{avg.}}_{\text{Nd-IO}} = 1.14$ and $\lambda^{\text{avg.}}_{\text{Ni}} = 0.37$) as shown in Fig. 3(d) by taking the average of $\lambda_{n\mathbf{k},m\mathbf{k}+\mathbf{q}}$ over all the $|m\mathbf{k}+\mathbf{q}\rangle$ states on the FS. Consequently, two-gap superconductivity arises in $Nd_{0.8}Sr_{0.2}NiO_2$ (Fig. 3(b)).

Solving the anisotropic Eliashberg equations [36-40] (with $\mu^* = 0.05$) at different temperatures yields superconducting gaps as a function of $T$ (Fig. 4(a)). We obtain a theoretical $T_c = 22.3$ K for $Nd_{0.8}Sr_{0.2}NiO_2$, agreeing well with the experimental $T_c = 19.3$ K at $x = 0.2$ [5], but is in strong contrast with the DFT result of $T_c = 0$ K [47]. Fig. 4(b) shows the computed superconducting quasiparticle DOS, $\frac{N_s(\omega)}{N_F}$, where the dominantly two-gap feature clearly shows up. We extract two "characteristic" superconducting gaps by defining the peak-to-peak energy separations (between corresponding peaks at positive and negative energies with respect to $E_F$) in Fig. 4(b) as $2\Delta$ (note the difference from the averaged gap values). By fitting their temperature dependence with $\Delta(T) = \Delta(T=0)\left[1-\left(\frac{T}{T_c}\right)^p\right]^{0.5}$ ($p$ is a fitting exponent) [40], we obtain zero-temperature gap values of $\Delta_1$ = 4.7 meV (with $p_1$ = 2.9) on the Nd-IO band and $\Delta_2$ = 2.1 meV (with $p_2$ = 2.3) on the Ni band.

Tunneling experiments can measure the superconducting quasiparticle DOS, where typically, an $s$-wave superconducting gap features a U-shape full-gap profile and a $d$-wave gap features a V-shape profile. Recent STS measurements [23] on $Nd_{0.8}Sr_{0.2}NiO_2$ thin films reveal both U-shape and V-shape profiles (and their mixture), casting mysteries on the pairing symmetry. However, measurements on infinite-layer nickelates suffer from relatively low sample quality originated from the chemical reduction procedure in the synthesis. Hence, we infer that surface spatial inhomogeneities could smear an intrinsic two-gap U-shape profile into a V-shape profile in the tunneling measurement, similar to the case of $MgB_2$ [97,98]. To demonstrate this effect, we introduce an extrinsic Gaussian broadening parameter $\sigma_{inhom}$ to the directly computed superconducting quasiparticle DOS and compare the smeared spectra with experiment [23]. With a small $\sigma_{inhom}$ = 0.1 meV (a clean and homogeneous surface), the theoretical curve nicely agrees with the experimental U-shape profile [23] (Fig. 4(c)). With a large $\sigma_{inhom}$ = 1.0 meV (a more defective surface), the peaks of the two $s$-wave gaps are significantly smeared, leading to a V-shape profile consistent with results from some parts of the samples in the experiment [23] (Fig. 4(d)). The different superconducting gap profiles from varying $\sigma_{inhom}$ also reproduce the experimentally observed superfluid density with both exponential and power-law behaviors as $T \rightarrow 0$ K, suggesting the importance of sample quality and providing insights for the experimental discrepancies [24,25]. (Supplemental Material [47] contains further analyses on STS and superfluid density results.)

With varying hole concentration $x$, the experimental phase diagram of $Nd_{1-x}Sr_xNiO_2$ presents a superconducting dome [3-5] existing in a hole doping range of $0.1 < x < 0.3$ [5]. Towards the underdoped regime with $x < 0.1$, there exists a charge ordering phase [99-101], which may compete with the superconducting phase (such as in $Ba_{1-x}K_xBiO_3$ [30,102,103]) and is beyond the scope of this work. For $x > 0.1$, we solve the anisotropic Eliashberg equations with $GW$ electronic structures at different doping levels, and obtain the doping-dependent $T_c$ of $Nd_{1-x}Sr_xNiO_2$ in Fig. 4(e). (Here we focus on the impact from the band structure and take the $e$-ph matrix elements and phonon dispersions as their values at $x$ = 0.2 [47].) A good agreement is obtained on the doping-dependent $T_c$ between the $GW$-Eliashberg results and experiments, while the remaining discrepancies may come from approximations and idealized crystal structures adopted in the theories. From our doping-dependent $GW$ band structure calculations (Fig. 4(f)), a non-rigid-band behavior is observed: the Nd-IO band and the Ni band respond to doping at different rates (a similar trend also exists in DFT results [45]). With increasing hole concentration $x$, the Nd-IO band with strong $e$-ph couplings moves up in energy away from $E_F$ (Fig. 4(f)), significantly reducing the number of Nd-IO states on the FS, thus leading to a rapid drop of the number of states with large $\lambda_{n\mathbf{k}}$ and the $e$-ph coupling constant $\lambda$ averaged over the Fermi surface (Fig. S10) and superconducting $T_c$ (Fig. 4(e)).

In conclusion, our *ab initio GW*-based anisotropic Eliashberg theory calculations and analyses show that within this theoretical framework, infinite-layer nickelate $Nd_{1-x}Sr_xNiO_2$ hosts a two-gap $s$-wave superconductivity (as in $MgB_2$ [40,96,97,104] and $CaC_6$ [105-107]), arising from strong self-energy effects on the two-band FS and $e$-ph couplings that are previously unforeseen. Note that this work has not explored other mechanisms (e.g., spin fluctuations [108], pairing via incipient bands [109,110]), which could possibly compete or cooperate with the $e$-ph coupling mechanism and may lead to exotic phenomena to be explored in the nickelates. Future STS and angle-resolved photoemission experiments [111] on samples with improved quality are expected to shed more light on the electronic structure and superconductivity of this class of materials.

**Acknowledgements**

This work was supported primarily by the Theory of Materials Program at the Lawrence Berkeley National Laboratory (LBNL) through the Office of Basic Energy Sciences, U.S. Department of Energy under Contract No. DE-AC02-05CH11231, which provided for the DFT, DFPT, *GW* and *GW*PT calculations. Advanced code for *GW* and *GW*PT calculations were provided by the Center for Computational Study of Excited-State Phenomena in Energy Materials (C2SEPEM) at LBNL, which is funded by the U.S. Department of Energy, Office of Science, Basic Energy Sciences, Materials Sciences and Engineering Division under Contract No. DE-AC02-05CH11231, as part of the Computational Materials Sciences Program. The development of the ABINIT-BerkeleyGW-EPW interface code was supported by the National Science Foundation under Grant No. OAC-2103991, allowing for solving $e$-ph coupling properties with *GW* band structures and *GW*PT $e$-ph matrix elements. The analysis on the doping dependence and superfluid density towards the end of this project was partially supported by the University of Southern California (Z.L.). Computational resources were provided by Frontera at TACC, which is supported by National Science Foundation under Grant No. OAC-1818253, and by Cori at National Energy Research Scientific Computing Center (NERSC), which is supported by the Office of Science of the U.S. Department of Energy under Contract No. DE-AC02-05CH11231. Z.L. thanks Yijun Yu for many fruitful discussions. The authors also thank Jiawei Ruan, Jingwei Jiang, and Qisi Wang for helpful discussions.

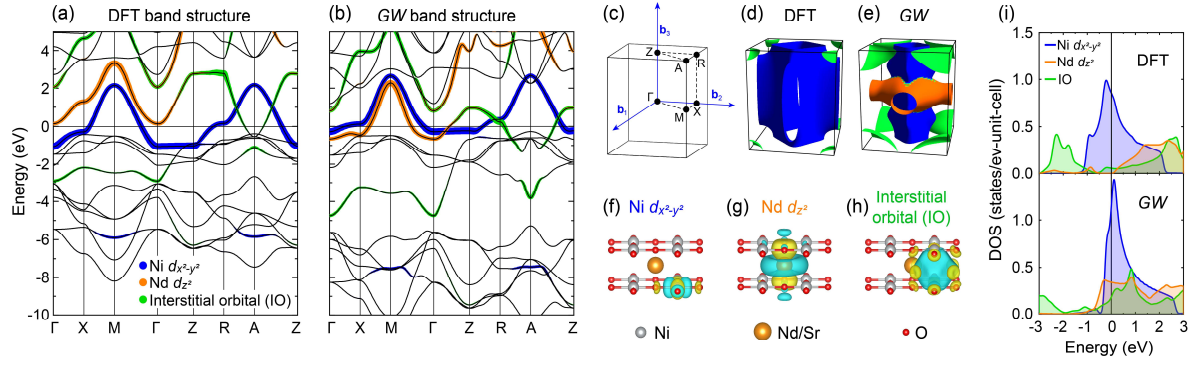

**Fig. 1.** (a) DFT and (b) *GW* band structure of Nd$_{0.8}$Sr$_{0.2}$NiO$_2$. (c) Brillouin zone. (d) DFT and (e) *GW* Fermi surface with orbital projections. (f-h) Three main basis MLWFs near $E_\text{F}$: (f) Ni $d_{x^2-y^2}$ (blue), (g) Nd $d_{z^2}$ (orange), and (h) interstitial orbital (green). Their projections are represented by colors on the band structure (proportional to the thickness of colored lines) in (a) and (b) and Fermi surface in (d) and (e). (i) Projected DOS from DFT and *GW* calculations.

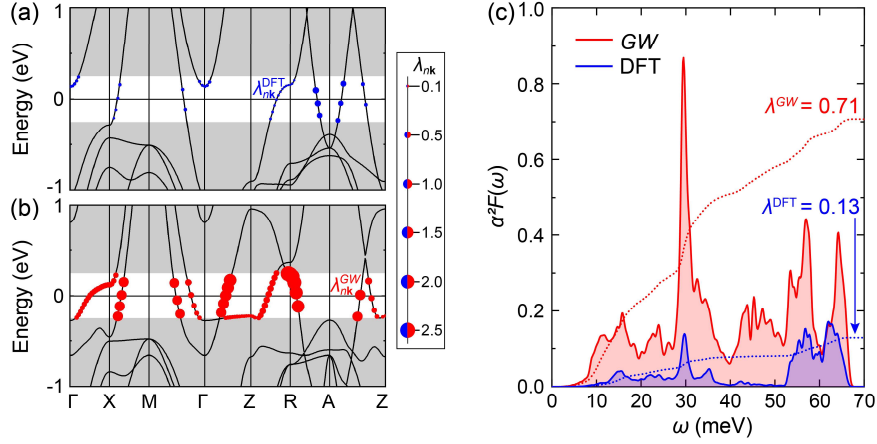

**Fig. 2.** State-resolved $e$-ph coupling strength $\lambda_{n\mathbf{k}}$ (for states within $E_F \pm 0.25$ eV) from (a) DFT and (b) $GW$ calculations. (c) Eliashberg function $\alpha^2 F(\omega) = \frac{1}{N_F}\Sigma_{mn\nu}\Sigma_{\mathbf{kq}}|g_{mn\nu}(\mathbf{k},\mathbf{q})|^2\delta(\varepsilon_{n\mathbf{k}} - E_F)\delta(\varepsilon_{m\mathbf{k+q}} - E_F)\delta(\hbar\omega - \hbar\omega_{\mathbf{q}\nu})$ (overall $e$-ph coupling strength as a function of phonon frequency) at the DFT (blue) and $GW$ (red) levels. The dotted lines represent the running integral $\lambda^<(\omega) = 2\int_0^\omega \frac{\alpha^2 F(\omega')}{\omega'}d\omega'$ which includes contributions from phonons with frequencies smaller than $\omega$. The standard $e$-ph coupling constant $\lambda = \lambda^<(\omega \to \infty)$ is 0.71 from the $GW$ calculation and is 0.13 from the DFT calculation.

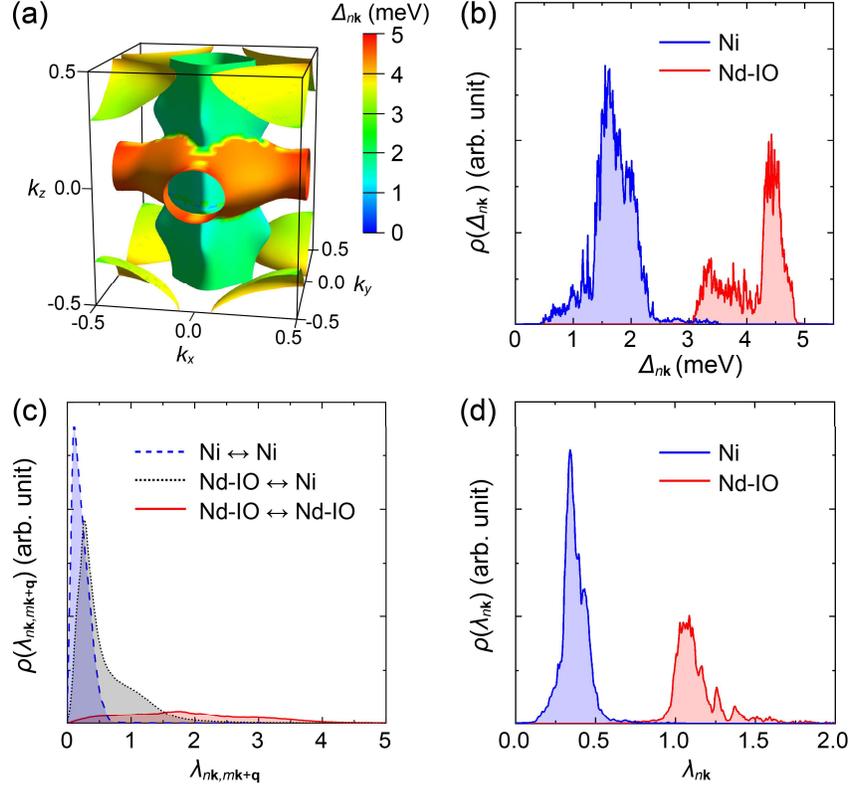

**Fig. 3.** (a) Value of the superconducting gap $\Delta_{n\mathbf{k}}$ (at $T = 5$ K) on the Fermi surface from $GW$-Eliashberg equations with $\mu^* = 0.05$. (b) Distribution of $\Delta_{n\mathbf{k}}$ for states on the Ni band and Nd-IO band. $\rho(x)$ is the density with value $x$. (c) Distribution of state-pair-resolved $e$-ph coupling strength $\lambda_{n\mathbf{k},m\mathbf{k+q}}$, decomposed to different transitions near $E_F$. (d) Distribution of state-resolved $e$-ph coupling $\lambda_{n\mathbf{k}}$ for states on the Ni and Nd-IO bands, respectively. Distributions shown in (b-d) are for states within $E_F \pm 0.25$ eV.

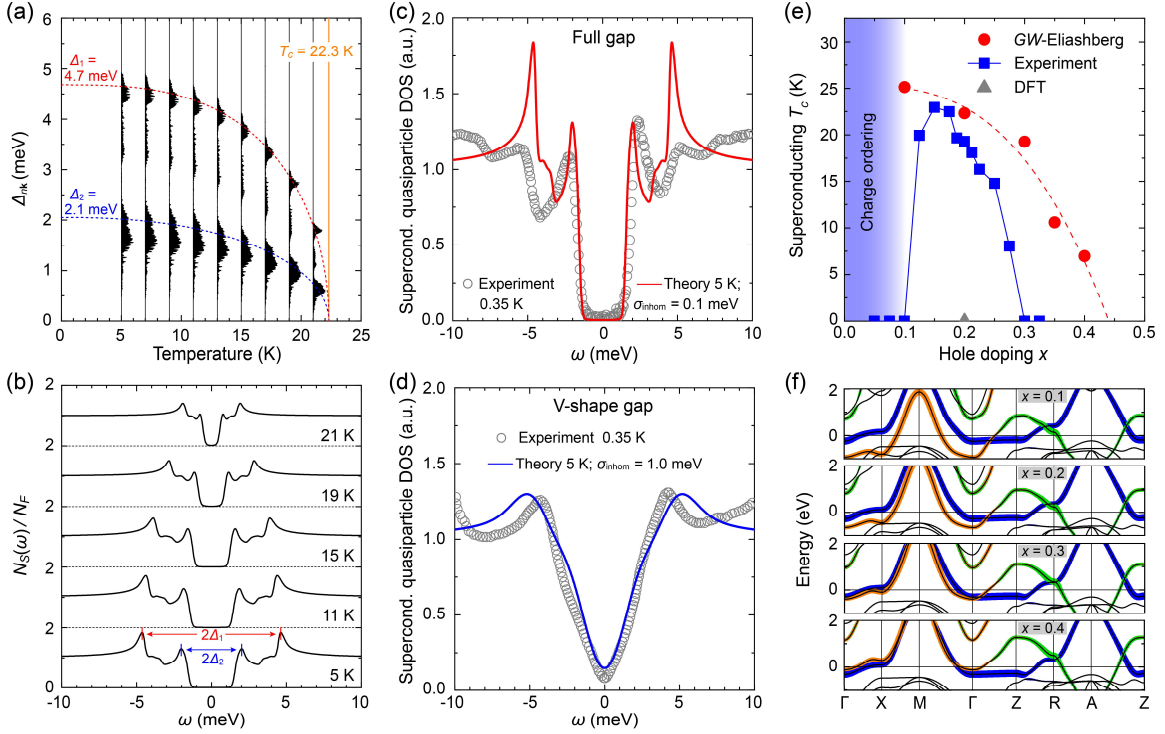

**Fig. 4.** (a) Distribution of $\Delta_{n\mathbf{k}}$ of $Nd_{0.8}Sr_{0.2}NiO_2$ at various temperatures. The dashed lines in (a) show the temperature dependence of the two characteristic gaps extracted from (b) the superconducting quasiparticle DOS, $\frac{N_S(\omega)}{N_F}$. (c, d) Gaussian-smeared superconducting quasiparticle DOS by a crystalline imperfection broadening parameter $\sigma_{\text{inhom}}$ comparing with experimental STS gap profiles [23]. (e) Superconducting $T_c$ as a function of hole doping concentration $x$ from theory and experiment [5]. The DFT $T_c$ is calculated using the McMillan-Allen-Dynes formula. (f) Doping-dependent $GW$ band structure of $Nd_{1-x}Sr_xNiO_2$ near $E_F$. The colors represent the orbital projections defined in Fig. 1.

*Supplemental Materials for*

# Two-Gap Superconductivity and Decisive Role of Rare-Earth *d* Electrons in Infinite-Layer Nickelates

Zhenglu Li[1,2,3] and Steven G. Louie[1,2*]

[1]*Department of Physics, University of California at Berkeley, Berkeley, CA, USA.*

[2]*Materials Sciences Division, Lawrence Berkeley National Laboratory, Berkeley, CA, USA.*

[3]*Mork Family Department of Chemical Engineering and Materials Science, University of Southern California, Los Angeles, CA, USA.*

*Email: sglouie@berkeley.edu

## 1. *GW* and *GW*PT methods

Density functional theory (DFT) is a popular computational approach for electronic structure [27,32,33] of materials. However, in its static version, it is a ground-state theory and the Kohn-Sham eigenvalues do not describe well the excited-state electronic quasiparticle properties including the band gap, bandwidth, and lifetime [28,29]. In many-body quantum theory, the many-electron correlation effects on quasiparticle excitations are casted into a quantity named self-energy $\Sigma(\mathbf{r}, \mathbf{r}'; \omega)$, which is non-local and frequency-dependent. Describing quasiparticle-related quantities (including band structure and *e*-ph coupling) using properties of the fictitious Kohn-Sham independent particles within DFT can be viewed as approximating the quasiparticle self-energy $\Sigma(\mathbf{r}, \mathbf{r}'; \omega)$ by the static and local DFT exchange-correlation potential $V_{\text{xc}}(\mathbf{r})$. Such an approximation sometimes can be drastic and lead to failure in describing materials with stronger electron correlations.

The *GW* approximation to the quasiparticle self-energy within the framework of many-body perturbation theory is a physically motivated and established method for quasiparticle excitations in solids from first principles [26,28,29]. Within the *GW* approximation [26,34,35], $\Sigma = iGW$ (*G* is the single-particle Green's function and *W* is the screened Coulomb interaction) and its non-locality and frequency-dependency are accurately captured for many materials. The *GW*-based methods have become standard *ab initio* computational approaches to describe the quasiparticle band structure as well as the optical excitations [28,29] (via the *GW*-Bethe-Salpeter equation method [29,48]) in studies of real materials. Recently, the *GW* approach has been extended to compute *e*-ph matrix elements via the *GW* perturbation theory (*GW*PT) approach [30,31], which includes the many-electron correlation effects to properly describe the quasiparticle *e*-ph coupling in materials.

In the *ab initio GW* method [26], in its simplest formulation, the quasiparticle excitation energy (e.g., band structure in crystals) is obtained by replacing the DFT $V_{xc}$ contribution to the Kohn-Sham particle by the *GW* self-energy contribution to the quasiparticle [26],

$$\varepsilon_{n\mathbf{k}}^{GW} = \varepsilon_{n\mathbf{k}}^{\text{DFT}} - \langle\psi_{n\mathbf{k}}|V_{xc}|\psi_{n\mathbf{k}}\rangle + \langle\psi_{n\mathbf{k}}|\Sigma|\psi_{n\mathbf{k}}\rangle.$$

Similarly, the *GW* level *e*-ph matrix element is obtained via *GW*PT as [30],

$$g_{mn\nu}^{GW}(\mathbf{k},\mathbf{q}) = g_{mn\nu}^{\text{DFT}}(\mathbf{k},\mathbf{q}) - \langle\psi_{m\mathbf{k}+\mathbf{q}}|\partial_{\mathbf{q}\nu}V_{xc}|\psi_{n\mathbf{k}}\rangle + \langle\psi_{m\mathbf{k}+\mathbf{q}}|\partial_{\mathbf{q}\nu}\Sigma|\psi_{n\mathbf{k}}\rangle,$$

where $\partial_{\mathbf{q}\nu}$ is a differential operator to phonon displacement for mode $\mathbf{q}\nu$ [30,31,33], and the DFT *e*-ph matrix elements are usually computed via the density-functional perturbation theory (DFPT) [32,33].

## 2. Computational details

Three major computational steps are carried out in this work. 1) DFPT calculations are performed with an in-house modified version of the ABINIT code [49]; 2) *GW*PT calculations are performed with the BerkeleyGW code (developers' version) [50,51]; and 3) Wannier interpolation of the *e*-ph matrix elements and the full **k**-dependent anisotropic Eliashberg theory calculations are performed with an in-house modified version of the EPW code [52]. An in-house developed wrapper code [53] prepares the band energies and the *e*-ph matrix elements (and other necessary quantities) from ABINIT (for DFT and DFPT), BerkeleyGW (for *GW* and *GW*PT), or in general any *ab initio* software packages, into files for the modified EPW to read. The Wannierization of the wavefunctions are done with the Wannier90 code [54]. Throughout the workflow, it is central that the gauge is kept the same among wavefunctions, Wannierization, phonon-induced first-order change in wavefunctions, and *e*-ph matrix elements at both DFPT and *GW*PT levels [53]. A consistent gauge is required for this workflow because the *e*-ph matrix elements and Wannierization basis transformation are gauge-dependent.

*DFT and DFPT calculations.* The DFT and DFPT calculations of $Nd_{1-x}Sr_xNiO_2$ are carried out using the ABINIT code [49]. There are typically two ways to computationally treat the Sr-induced charge doping effects within a primitive unit cell: 1) the virtual crystal approximation (VCA) [55] where the pseudopotentials of Nd and Sr are linearly interpolated (i.e., mixed) depending on *x*; 2) by changing the number of electrons of the system and adding a uniform background charge to avoid charge imbalance. We have tested both approaches by relaxing the *c* lattice vector of the bulk crystal structure of $Nd_{1-x}Sr_xNiO_2$, while keeping the in-plane lattice constant fixed at *a* = 3.905 Å to simulating the $SrTiO_3$ substrate effects [2]. Fig. S1 shows the optimized length of the *c* lattice vector as a function of *x* (hole doping), and its comparison to experiment [3]. VCA nicely agree with experiment in the *c* lattice vector length and the doping trend, whereas the background compensating charge method shows an opposite trend. Therefore, we adopt the VCA approach for the Sr doping effect. For $Nd_{0.8}Sr_{0.2}NiO_2$ (i.e., $x = 0.2$), the VCA-relaxed $c = 3.3375$ Å is used throughout the calculations.

For the DFT and DFPT calculations, we use norm-conserving scalar-relativistic pseudopotentials (giving totally 40.8 valence electrons per $Nd_{0.8}Sr_{0.2}NiO_2$ formula unit) taken from the PseudoDojo library [56]. The rare-earth element Nd contains strongly localized *f* electrons, which are shown to be few eV away from the Fermi energy in all-electron DFT+*U* calculations [57]. Therefore, in this work, we use a Nd pseudopotential where the *f* electrons are frozen as core electrons, which has been shown to be a standard best practice treatment [56]. The exchange-correlation functional used is the generalized gradient approximation (GGA) developed by Perdew, Burke, and Ernzerhof [58]. A planewave cutoff of 100 Ry is used for wavefunctions. The electron and phonon Brillouin zones (BZs) are sampled by a 12×12×12 **k**-grid (DFT and DFPT) and a 6×6×6 **q**-grid (DFPT), respectively.

*Magnetism analysis.* Experimentally, no long-range magnetic order is observed in infinite-layer nickelates [2], and signatures of short-range antiferromagnetic correlations between local Ni moments have been revealed by resonant inelastic x-ray scattering (RIXS) [59] and muon spin rotation/relaxation experiments [60]. Early DFT-based theoretical studies [61-64] have discussed changes in the electronic structure due to magnetism, typically in the long-range ferromagnetic or antiferromagnetic orders, instead of short-range fluctuations. Here, to study the effects of the local antiferromagnetic spin fluctuations on the electronic structure (globally paramagnetic), we perform explicit spin-polarized DFT calculations to approximately simulate this phase [65]. We focus on the undoped $NdNiO_2$ which shows the strongest RIXS spectral features [59]. We set up a 4×4×4 supercell that contains 64 Ni sites. A *G*-type antiferromagnetic order is initialized, and the spins of 20 randomly picked sites are flipped to create global paramagnetism (total magnetization is still zero). The short-range antiferromagnetic correlation is simulated such that the number of nearest-Ni-neighbor (both in-plane and out-of-plane) antiferromagnetic pairs ($N_{pair}^{AF}$) is always larger than that of the ferromagnetic pairs ($N_{pair}^{FM}$). We randomly generate four spin configurations, with their $\frac{N_{pair}^{AF}}{N_{pair}^{FM}} = \frac{54.2\%}{45.8\%}, \frac{55.2\%}{44.8\%}, \frac{55.2\%}{44.8\%}$, and $\frac{58.3\%}{41.7\%}$, respectively. A representative supercell spin configuration is shown in Fig. S2(a). We compute the band structures of the magnetic supercells and unfold [66,67] them into the Brillouin zone of the primitive unit cell as shown in Fig. S2(b), averaging all four spin configurations. Near $E_F$, the unfolded Nd-IO spectral intensity closely traces the nonmagnetic DFT Nd-IO band. Meanwhile, the Ni spectral intensity is broadened due to the magnetic configurations, resembling the spin fluctuation behavior to some extent. Our results suggest that in the paramagnetic phase of infinite-layer nickelates with short-range antiferromagnetic correlations, the Nd-IO band remains nearly intact by the spin structures, and the Ni band still shows considerable spectral intensity around $E_F$ (but more scattered), compared with the non-magnetic DFT band structure. Since the superconductivity is mainly driven by the *e*-ph coupling from the Nd-IO band, for the calculations and analysis in this work, the non-spin-polarized electronic structure is a good approximation. Therefore, we neglect the effects of spin polarization in our main results.

*Wannierization.* Maximally localized Wannier functions (MLWFs) [43] are often used to analyze the electronic structures of infinite-layer nickelates [11,13, 42,68-72]. In this work, we use MLWFs for the analysis of electronic structure [43] and interpolation of *e*-ph matrix elements. The Wannierization of the DFT wavefunctions are performed with the Wannier90 package [54]. We use a subspace of 17 MLWFs to capture the multi-orbital nature of nickelates and to interpolate to a highly dense grid of **k**-points of the electronic structure accurately. For the initial projections, we use five *d* orbitals on Nd atom, five *d* orbitals on Ni atom, three *p* orbitals for each O atom, and one *s* orbital on the interstitial hollow site on the Nd plane. The optimized MLWFs maintain the major characters of the initial projections. The Wannierization is done with a 6×6×6 **k**-grid. This set of MLWFs are used to interpolate the DFT and *GW* band structures, and the DFPT and *GW*PT *e*-ph matrix elements. The *GW* Fermi energy $E_F$ is determined with Wannier-interpolated DOS and electron counting. The Fermi surfaces in Figs. 1(d), 1(e) and 3(a) are plotted using the software FermiSurfer [73]. The MLWFs in Figs. 1(f-h) are plotted using the software VESTA [74], and the iso-surface level is set by default as $d_{iso} = \langle |w(\mathbf{r})| \rangle + 2\sigma(|w(\mathbf{r})|)$, where $w(\mathbf{r})$ is the volumetric data of a MLWF, and $\langle |w(\mathbf{r})| \rangle$ and $\sigma(|w(\mathbf{r})|)$ are the average value and standard deviation of the absolute values of $w(\mathbf{r})$. The localization of MLWFs can be quantified by their spreads [43]. In particular for the dominant characters near $E_F$, the spreads are 0.455 Å$^2$ for Ni $d_{x^2-y^2}$ orbital, 2.814 Å$^2$ for Nd $d_{z^2}$ orbital, and 3.002 Å$^2$ for the interstitial orbital (IO).

*GW and GWPT calculations.* The *GW* and *GW*PT calculations are performed with the BerkeleyGW package [51]. The inverse dielectric matrix $\epsilon^{-1}(\mathbf{p})$ is computed with 200 bands and a screened Coulomb cutoff of 25 Ry on a 12×12×12 **p**-grid with a 12×12×12 **k**-grid sampling. The $|\mathbf{p}| \to 0$ point is represented by $\mathbf{p}_0 = (\frac{1}{48}, 0, 0)$ and the $\epsilon^{-1}(\mathbf{p}_0)$ is computed with a 48×48×48 **k**-grid sampling with 28 bands where intra-band transitions dominate. The Hybertsen-Louie generalized plasmon-pole model [26] is used for the frequency dependence in $\epsilon^{-1}$. The construction of the self-energy operator $\Sigma$ and its phonon-induced first-order change $\partial_{\mathbf{q}\nu}\Sigma$ uses a bare Coulomb cutoff of 60 Ry and a screened Coulomb cutoff of 25 Ry, along with 200 bands. Our *GW* electronic structure is consistent with the experimentally revealed electronic structure [75] (where the O states are mainly lying below the occupied Ni states) and hybridized multi-character states [76-78] near $E_F$. The most computationally intensive step is the *GW*PT calculation, which is 2 – 3 orders of magnitude more expensive than a standard *GW* quasiparticle calculation.

All main results for the DFT and DFPT parts presented in this work are carried out using ABINIT and VCA (unless otherwise stated), and they are used as the starting point for the *GW* and *GW*PT calculations (ABINIT-BerkeleyGW). We have validated our *GW* calculations using another independent electronic code Quantum Espresso (QE) [79] to perform the DFT starting-point calculations for the non-doped compound NdNiO$_2$ (as well as the magnetism analysis and Fig. S2). A similar *GW* band renormalization (i.e., with significantly more Nd-IO characters appears near $E_F$) is seen using QE-BerkeleyGW. For the *GW* convergence tests, we have checked with 500 and 1000 bands, a bare Coulomb cutoff of 100 Ry, and eigenvalue self-consistent *GW* iterations (up to $G_3W_0$) in constructing the self-energy operator, and the resulting band structures remain qualitatively similar. Higher convergence criteria are prohibitive for *GW*PT calculations, but we demonstrate below that the current moderate convergence captures the main physics and the conclusions are robust (see discussions below on numerical/convergent uncertainties). In this work, we present the $G_0W_0$ band structure and the $G_0W_0$PT *e*-ph calculations with the same convergence parameters to keep the level of theory consistent. We do not update the wavefunctions in the *GW* calculations as the two bands near $E_F$ have quite distinct characters and do not mix much. In some parts of the band structure (e.g., between –1 to –4 eV at the A point in Fig. 1(b)), unusual dispersions arise, indicating the wavefunctions might need to be updated for these states [80,81]. However, because these states are away from $E_F$, we anticipate their effects on the *e*-ph coupling for states near $E_F$ and superconductivity properties are small. Using *GW*-updated wavefunctions for DFPT and *GW*PT *e*-ph matrix elements calculations (which is unnecessary for this work) is a grand challenge at the current stage of the theory and method development in the community.

*Uncertainties in GW calculations.* Since there are no available angle-resolved photoemission measurements on the band structures as a reference, here we estimate the uncertainties in the *GW* calculations [82] and discuss the robustness of our conclusions. Fig. S3(a) shows the band structures of undoped NdNiO$_2$ computed with various standard *GW* schemes using fully converged parameters (100 Ry of bare Coulomb cutoff and 1000 bands in self-energy evaluations). Eigenvalue self-consistency enhances the self-energy effects found in the $G_0W_0$ calculations, i.e., moving the Nd-IO band further down in energy, leading to stronger *e*-ph coupling and higher $T_c$ (assuming unchanged *e*-ph matrix elements). In this work, we use the $G_0W_0$ on-shell evaluation of the self-energy (i.e., evaluation at DFT energies) because, i) the resulting band structure is closer to the eigenvalue self-consistent *GW* calculation compared with off-shell evaluation, and ii) the level of approximation is consistent with our $G_0W_0$PT implementation and calculations. The variations in Fig. S3(a) represent the uncertainties of the *GW* calculations in nickelates, which does not alter the qualitative electronic properties. Fig. S3(b) shows the fully converged (100 Ry/1000 bands) $G_0W_0$ band structure of Nd$_{0.8}$Sr$_{0.2}$NiO$_2$, showing that our adopted convergence (60 Ry/200 bands in Fig. 1(b)) is comparable to the expected uncertainties and reasonably capture the main electronic feature of the increased Nd-IO states on $E_F$ comparing with the DFT result in Fig. 1(a).

To quantify the effects from the expected uncertainties and convergence in superconductivity, we perform *GW*-Eliashberg calculations using the fully converged $G_0W_0$ band structures (100 Ry/1000 bands) at $x = 0.0$ and $x = 0.2$ (see doping dependence discussions below for technical details). At $x = 0.0$, we obtain the *GW*-Eliashberg $T_c = 18.5$ K, and at $x = 0.2$, we obtain the *GW*-

Eliashberg $T_c$ = 12.7 K, showing robust superconductivity comparable to experiments. Note that eigenvalue self-consistency iterations will further enhance $T_c$ due to further increased Nd-IO characters at $E_F$ (Fig. S3(a)). Calculations with higher convergence stringency and eigenvalue self-consistency may lead to a better agreement with experiments in $T_c$ compared with the reported results (Fig. 4(e)). However, the comparison between theory and experiments is still subject to other uncertainty sources, including the approximations adopted, the value of $\mu^*$, and the experimental determination of $T_c$. Therefore, we do not expect or aim at a perfect agreement but emphasize on the qualitative behavior (such as *s*-wave two-gap superconductivity) and trend. Note that the quantitative numerical variations from the uncertainties in *GW* calculations and the convergence threshold we adopted can alternatively be viewed as uncertainties in the doping axis. In the limit of undoped parent compound NdNiO$_2$, we obtain strong *e*-ph coupling and superconductivity, which set up the robust high-$T_c$ superconductivity and the half-dome structure towards the high hole-doping regime. Our analyses arrive at the same physical conclusion of a phonon-mediated two-gap *s*-wave superconductivity in hole-doped Nd$_{1-x}$Sr$_x$NiO$_2$. Future high-quality angle-resolved photoemission data could provide critical information in resolving the band structure of infinite-layer nickelates.

*EPW calculations.* We use the EPW software package [52] to carry out the Wannier interpolation of the *e*-ph matrix elements and the **k**-dependent anisotropic Eliashberg theory calculations [37-39]. The DFPT is used to compute *e*-ph matrix elements $g_{mn\nu}^{\text{DFT}}(\mathbf{k}, \mathbf{q})$ across all needed bands covering the Wannier subspace, whereas *GW*PT is employed to compute *e*-ph matrix elements $g_{mn\nu}^{GW}(\mathbf{k}, \mathbf{q})$ on the two bands crossing the Fermi energy (due to computational expenses and also because only these two bands are most relevant to superconductivity), which has been shown to adequately capture the *GW* self-energy effects for properties near the Fermi surface [31]. The *e*-ph matrix elements on other bands not computed by *GW*PT take their DFPT values. The *e*-ph matrix elements from both DFPT and *GW*PT are directly computed on 6×6×6 **k**-grid and **q**-grid, and then interpolated by Wannier functions. Fig. 2(b) and Fig. 3(c) use interpolated 36×36×36 **k**-grid and 18×18×18 **q**-grid with an electron smearing of 0.05 eV and a phonon smearing of 0.5 meV. All other presented anisotropic Eliashberg quantities are solved on finer 40×40×40 **k**-grid and **q**-grid for better resolution and visualization (Fig. 3(a) has been further numerically interpolated to 80×80×80 **k**-grid). These two sets of grids are both converged, where the difference in total $\lambda$ is less than 2%. At different temperatures, the anisotropic Eliashberg equations are solved [52] along the imaginary frequency axis up to a cutoff of 0.5 eV. Padé approximants are used in the analytical continuation to the real frequency axis when needed [52].

## 3. Symmetry unfolding of *e*-ph matrix elements with gauge recovering

Due to the heavy computation expense of *GW*PT in computing $g_{mn\nu}^{GW}(\mathbf{k}, \mathbf{q})$, symmetry reduction in the number of matrix elements is desired. However, in combining with Wannier interpolation, an important issue arises. The Wannierization procedure for a set of wavefunctions on a full **k**-grid generates a basis transformation matrix with its gauge fixed to this unique set of wavefunctions. In performing the symmetry unfolding of wavefunctions and *e*-ph matrix elements, the unfolding process generates a different gauge at the unfolded **k**-point. The gauge must be recovered to the original and unique gauge of the full **k**-grid wavefunctions in order to continue with the Wannier interpolation procedure [53]. In another word, the symmetry-unfolded *e*-ph matrix elements (complex numbers) must be exactly the same (up to numerical convergence accuracy) as those directly computed without using any symmetries, for both their magnitude and the phase.

To achieve this goal, we developed and implemented a symmetry unfolding method for *e*-ph matrix elements with gauge recovering [53]. This symmetry unfolding technique is implemented in the BerkeleyGW code and is suitable for both DFPT and *GW*PT *e*-ph matrix elements. It is designed to request the direct computation of *e*-ph matrix elements only within the symmetry-reduced **q**-grid (along with the full **k**-grid for each **q**), and then full *e*-ph matrix elements on full **q**-grid (and full **k**-grid for each **q**) can be obtained by symmetry unfolding. The gauge of the unfolded matrix elements is recovered to be consistent with the original and unique set of wavefunctions, as if the *e*-ph matrix elements on full **q**- and **k**-grids were computed directly without using symmetry reduction.

## 4. Further analysis of *e*-ph coupling properties of infinite-layer nickelates

*Phonon properties.* We compute phonon properties using DFPT [32]. This is justified since the phonon frequency within the adiabatic approximation is obtainable from the DFT electronic ground-state total energy. The DFPT phonon frequencies are generally in good agreement with experiments [32] because static DFT is a ground-state theory. This is different from *e*-ph coupling, which is the interaction between a phonon and a quasiparticle in an excited state [30].

Fig. S4(a) shows the calculated phonon band structure, and Fig. S4(b) shows the total phonon density of states (DOS) and the partial phonon DOS decomposed to different atomic contributions. Comparing with the Eliashberg function $\alpha^2 F$ in Fig. 2(c) in the main text, we note that phonons with full range of frequencies and vibrations from all three atoms (Nd, Ni, and O) contribute to the *e*-ph coupling, among which the O vibrations play a dominant role. In Fig. S5, we disentangle the *GW* effects in the band structure (i.e., changes in the character of states crossing $E_F$) and those in the *e*-ph matrix elements on the resulting $\alpha^2 F$ spectrum. Using the DFT band structure for the electronic states, in Fig. S5(a), we show the computed $\alpha^2 F$ using either the DFPT or the *GW*PT *e*-ph matrix elements, and obtain $\lambda(\varepsilon^{\mathrm{DFT}}, g^{\mathrm{DFT}}) = 0.13$ and $\lambda(\varepsilon^{\mathrm{DFT}}, g^{GW}) = 0.21$. Using the *GW* band structure for the electronic states, in Fig. S5(b), and similarly using either the DFPT or the *GW*PT *e*-ph matrix elements, we obtain $\lambda(\varepsilon^{GW}, g^{\mathrm{DFT}}) = 0.48$ and $\lambda(\varepsilon^{GW}, g^{GW}) = 0.71$. Comparing the two blue curves in Fig. S5, the *GW* effects in the band states crossing $E_F$ induce an enhancement in $\lambda$ by a factor of 3.7 due to wavefunction character changes. Comparing the red and blue curves in Fig. S5(b), *GW*PT further enhances $\lambda$ by a factor of 1.5 due to correlation (self-energy) effects in the *e*-ph matrix elements. The overall enhancement factor of $\lambda^{GW}/\lambda^{DFT} = 5.5$ (Fig. 2(c)) thus comes from the combined *GW* renormalization effects in both the band characters and *e*-ph matrix elements near $E_F$.

*Decomposition analysis of e-ph coupling.* In Fig. 3, we have decomposed at the *GW*-level the superconducting gap $\Delta_{n\mathbf{k}}$, the state-resolved *e*-ph coupling $\lambda_{n\mathbf{k}}$, and the state-pair-resolved *e*-ph coupling $\lambda_{n\mathbf{k},m\mathbf{k}+\mathbf{q}}$ into contributions from the Ni and Nd-IO bands. As the electronic structure near $E_F$ from DFT significantly differs from *GW*, here we also present a similar analysis on the DFT *e*-ph coupling. Due to the very weak DFT *e*-ph coupling strength ($\lambda^{DFT} = 0.13$), it is extremely numerically demanding to solve for the superconducting gaps from the anisotropic Eliashberg equations (temperature needs to be set extremely low) and the corresponding superconducting gaps will be basically zero for any reasonable value of $\mu^*$. However, we can still analyze $\lambda_{n\mathbf{k}}$ and $\lambda_{n\mathbf{k},m\mathbf{k}+\mathbf{q}}$, as shown in Fig. S6. In Fig. S6(a), the Ni ↔ Ni couplings dominate at the DFT level. There is a small portion of Nd-IO ↔ Ni couplings, and a negligible portion of Nd-IO ↔ Nd-IO couplings. These distributions of the state-pair-resolved $\lambda_{n\mathbf{k},m\mathbf{k}+\mathbf{q}}$ lead to a Ni band dominant state-resolved *e*-ph coupling $\lambda_{n\mathbf{k}}$ as shown in Fig. S5**b**, but with very low coupling values. The overall DFT *e*-ph coupling strength is very small with $\lambda^{DFT} = 0.13$.

*Superconducting transition temperature $T_c$.* The superconducting $T_c$ is obtained by solving the anisotropic Eliashberg equations at different temperatures and then fitting the average gap using $\Delta_0(T) = \Delta_0(T=0)\left[1 - \left(\frac{T}{T_c}\right)^p\right]^{0.5}$, with fitted $\Delta_0(0) = 3.1$ meV, $p = 2.5$, and $T_c = 22.3$ K in Fig. 4(a) in the main text. In solving for superconductivity properties, along with the attractive interaction (in this work from the *e*-ph coupling), the residual Coulomb repulsion between electrons also needs to be considered [33,37]. In principle, the effective Coulomb interaction for states near the Fermi surface can be computed from first principles. Here, we use the standard alternative method to treat the Coulomb interaction by an effective parameter $\mu^*$ [33,37-40]. Figure S7 plots the anisotropic Eliashberg $T_c$ (at the *GW* level) with three physically reasonable $\mu^*$ values (0.05, 0.08, and 0.1), giving $T_c$ around 20 K consistent with experiments. Varying the full range of $\mu^*$ from 0.05 – 0.1 yields a change of $T_c$ only by 3 K. All results in the main text (Fig. 3 and Fig. 4) are reported with $\mu^* = 0.05$. Besides the anisotropic Eliashberg theory, we have also computed $T_c$ using the semi-empirical McMillan-Allen-Dynes formula [83,84] with our computed overall *e*-ph coupling constant $\lambda$ (usually only a good estimation for isotropic superconductors, i.e., $\lambda_{n\mathbf{k}}$ is independent of *n* and **k**), via $k_B T_c = \frac{\hbar\omega_{\log}}{1.2}\exp\left[-\frac{1.04(1+\lambda)}{\lambda-\mu^*(1+0.62\lambda)}\right]$, where $\omega_{\log}$ is the logarithmic average of the phonon frequencies. The McMillan-Allen-Dynes $T_c$ shows a large discrepancy and underestimation compared with the full **k**-dependent anisotropic Eliashberg theory. This highlights the necessity of using the anisotropic Eliashberg theory to capture the multi-gap nature of superconductivity in $Nd_{1-x}Sr_xNiO_2$. With the DFT-computed *e*-ph coupling strengths, the McMillan-Allen-Dynes formula gives a value of $T_c = 0$ K, and the anisotropic Eliashberg equations are not practically solvable accurately due to the very weak DFT coupling strength.

Our *GW*-based results reveal a strong *e*-ph coupling and phonon-mediated two-gap superconductivity in the infinite-layer nickelates, which provide some insights to recent magnetotransport experiments where an apparent violation of the Pauli limit ($H_P$) is observed in La- and Pr-nickelates [85-87] (but not in Nd-nickelates [88]). While previous analyses of the experiments adopt the weak-coupling value of $H_P = 1.84T_c$ [Tesla/Kelvin], our calculations show a strong *e*-ph coupling regime, which enhances $H_P$ by a factor [89,90] of ~ $(1+\lambda)$, eliminating or mitigating the violation. The effects of two-gap superconductivity [89] and experimental sample quality [85,87,89] (along with uncertainties in the measured $T_c$) further affect the measured upper critical field and the appropriate value of the Pauli limit.

*Doping dependence.* Fig. 4(e) in the main text presents the calculated doping-dependent *GW*-Eliashberg superconducting $T_c$. To capture the non-rigid-band behavior, at each doping level ($x = 0.1, 0.2, 0.3, 0.35, 0.4$), we perform *GW* calculations to obtain the band structures as shown in Fig. 4(f) in the main text. The qualitative behavior of the drop of $T_c$ with increasing hole doping should mostly come from the Nd-IO band movement, and other ingredients such as *e*-ph matrix elements and phonon frequencies likely only induce quantitative modifications. Therefore, to reduce the computational cost (especially for *GW*PT), we neglect the doping dependence in the *e*-ph matrix elements and phonon frequencies, but use the band structure at the actual doping levels. In the doping-dependent *GW*-Eliashberg calculations, the wavefunction gauge needs to be consistent, and therefore we assume the wavefunction characters remain unchanged. For each electronic state from $x = 0.2$, we assign its new band energy at different doping levels by finding the corresponding state with the maximum wavefunction overlap (calculated in the Wannier basis). This approach well captures the doping dependence in the electronic structure and its impact on the superconductivity, as shown in Fig. 4(e) in the main text.

*Superconducting quasiparticle DOS.* The superconducting quasiparticle DOS can be obtained from the gap solutions of the anisotropic Eliashberg equations [91],

$$\frac{N_S(\omega)}{N_F} = \sum_{n\mathbf{k}} \frac{\delta(\varepsilon_{n\mathbf{k}} - E_F)}{N_F} \text{Re}\left[\frac{(\omega + i\Gamma)}{\sqrt{(\omega + i\Gamma)^2 - \Delta_{n\mathbf{k}}^2(\omega)}}\right],$$

where $\Gamma$ is an intrinsic superconducting quasiparticle lifetime and we take $\Gamma = 0.01$ meV. The directly computed superconducting DOS is shown as dots in Fig. S8(a). In the main text and Fig. 4(c) and 4(d), we have introduced an additional extrinsic crystalline imperfection Gaussian broadening parameter $\sigma_{\text{inhom}}$ to the DOS. Fig. S8(a) shows the Gaussian smearing of the directly computed superconducting DOS with a small (0.1 meV) and a large (1 meV) broadening parameter. The superconducting quasiparticle DOS with $\sigma_{\text{inhom}} = 0.1$ meV here, as well as in Fig. 4(c) of the main text, reflects all the sharp features in the directly calculated intrinsic superconducting gap distribution. Fig. 4(b) in the main text correspond to results convoluted with $\sigma_{inhom}$ = 0.1 meV as well. Fig. S8(b) shows that the experimentally observed mixed U-shape and V-shape features [23] can be reproduced with a moderate $\sigma_{inhom}$ = 0.8 meV.

The experimental tunneling data in Fig. 4(c) and 4(d) in the main text and Fig. S8(b) are extracted from the bottom curve in Fig. 3**b** of Ref. [23], the bottom curve of Fig. 2**d** of Ref. [23], and Supplementary Fig. 6**a** of Ref. [23], respectively.

*Superfluid density.* The temperature dependence of the superfluid density $\rho_s$ reflects the superconducting pairing symmetry especially in the low-temperature regime. In the clean limit, as $T \rightarrow 0$ K, an *s*-wave full gap leads to an exponential saturation in $\rho_s$, and a nodal *d*-wave gap leads to a linear temperature dependence [92]. In the presence of strong scattering, the profiles of $\rho_s$ for both cases reduce to a power-law dependence in *T*. Recent experiments [24,25] revealed both exponential and power-law temperature dependence of $\rho_s$ in different infinite-layer nickelate thin films, and different pairing symmetries were proposed based on data fitting. Theoretically, the superfluid density is largely determined by the superconducting quasiparticle DOS. Our computed *s*-wave gaps are bimodally distributed on the different FS sheets, and if we neglect the variation of the gaps and Fermi velocities on each FS sheet [93], the superfluid density $\rho_s$ can be approximately written as $\rho_s(T) = 1 - \frac{2}{k_B T}\int_0^\infty \frac{N_s(\omega)}{N_F} f(\omega)[1-f(\omega)]d\omega$, where $f(\omega)$ is the Fermi-Dirac function. As discussed above, introducing different values of broadening parameter $\sigma_{inhom}$ gives rise to different profiles of $\frac{N_s(\omega)}{N_F}$. Fig. S9(a) and S9(b) show the normalized superfluid density $\frac{\rho_s(T)}{\rho_s(0)}$ computed with superconducting quasiparticle DOS profiles with $\sigma_{inhom}$ = 0.1 meV (clean limit) and 1.0 meV (strong scattering limit), respectively. The fine resolution of the temperature dependence (for $T/T_c \leq 0.3$) is achieved by extrapolating the $\frac{N_s(\omega)}{N_F}$ results at $T$ = 5 K with the fitted temperature dependence of the average gap $\Delta_0(T)$. Superfluid density with both exponential and power-law behaviors can arise within the two-gap *s*-wave superconductivity picture. Our *ab initio* results and analysis provide a predictive guidance to future measurements on nickelate samples with improved quality.

The experimental superfluid density data in Fig. S9(a) and S9(b) are extracted from Fig. 3 of Ref. [24] and Fig. 4 of Ref. [25].

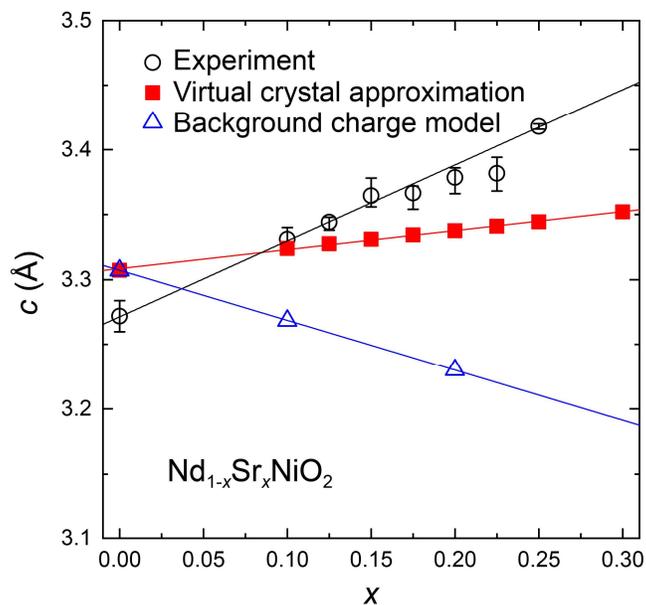

**Fig. S1.** Doping-dependent $c$ lattice vector length computed using the virtual crystal approximation (VCA) (red squares) and by an approximation which adds a uniform background charge and changes electron occupations (blue triangles). VCA shows better agreement with experiment [3] (black circles), and is thus used to compute all other results presented in this work.

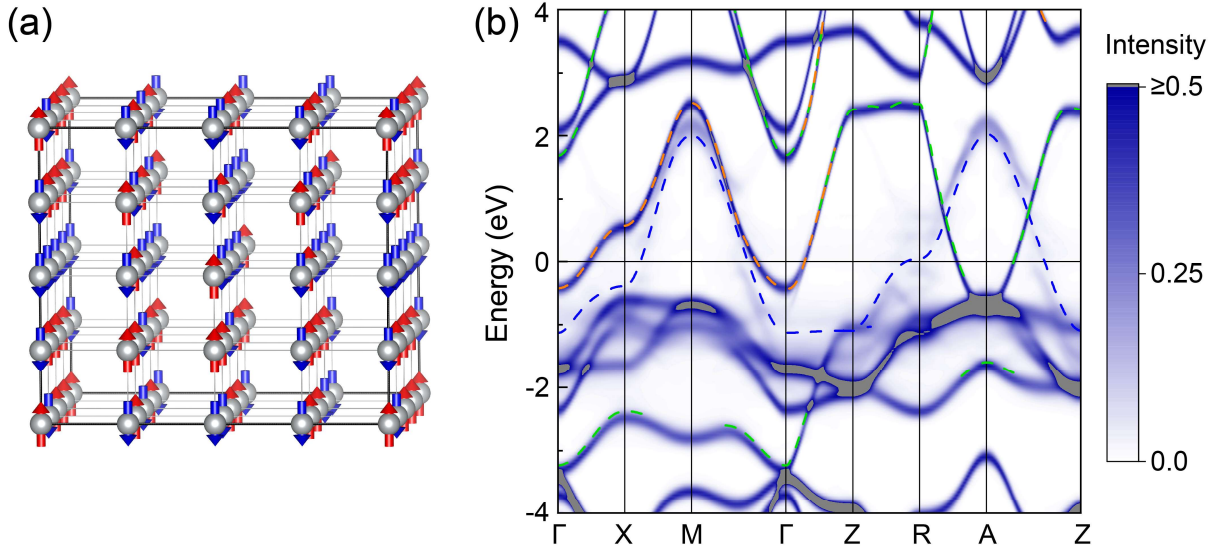

**Fig. S2.** (a) A representative paramagnetic spin configuration of NdNiO$_2$ with local antiferromagnetic correlations in a 4×4×4 supercell. Only Ni atoms are shown. (b) Unfolded spectral function or band structure (from spin-polarized DFT calculations) of paramagnetic phase with antiferromagnetic correlations plotted in the Brillouin zone of NdNiO$_2$ primitive unit cell. Four randomized magnetic configurations (with one shown in (a)) are included and averaged. The maximum projection intensity in the data is 1.0 and we cut off intensity higher than 0.5 (shown in gray color) which is typically due to band degeneracies. The dashed lines represent the band structure from nonmagnetic (non-spin-polarized) DFT calculations and the colors correspond to orbital projections to the Ni $d_{x^2-y^2}$ orbital (blue), Nd $d_{z^2}$ (orange), and IO (green) states. The Nd-IO band remains largely unaffected by the magnetic fluctuations, and the Ni-band shows spectral weight broadening near $E_F$ due to spin fluctuations.

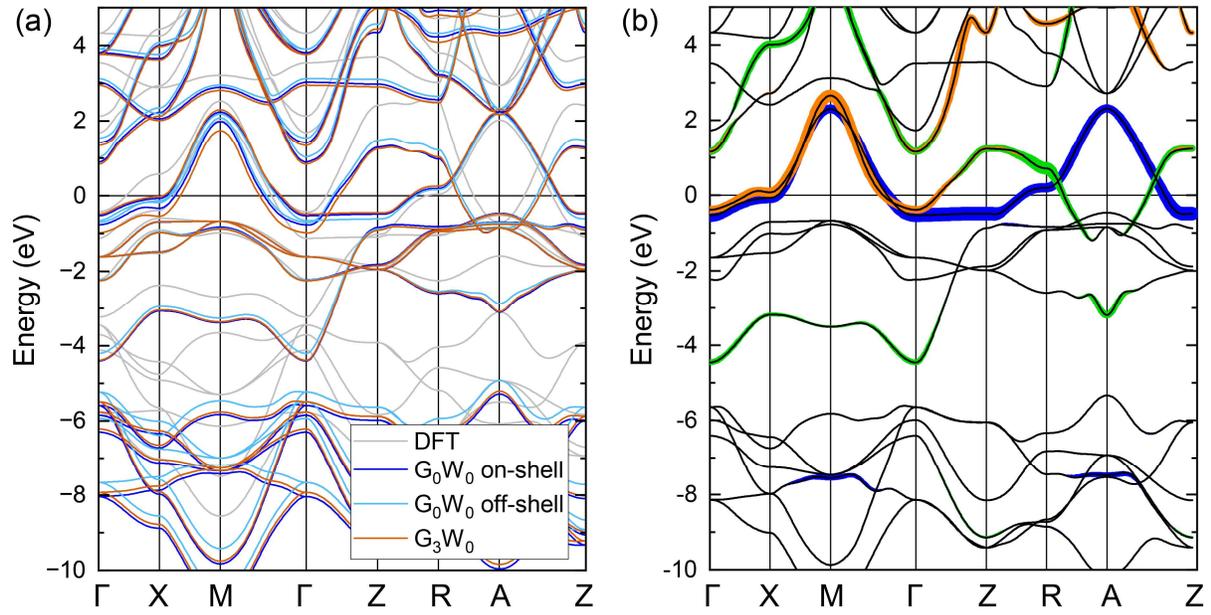

**Fig. S3.** (a) Band structures of undoped $NdNiO_2$ from DFT and different *GW* calculation schemes. (b) Band structures of $Nd_{0.8}Sr_{0.2}NiO_2$ from $G_0W_0$ calculations with 100 Ry bare Coulomb cutoff and 1000 bands in self-energy summation.

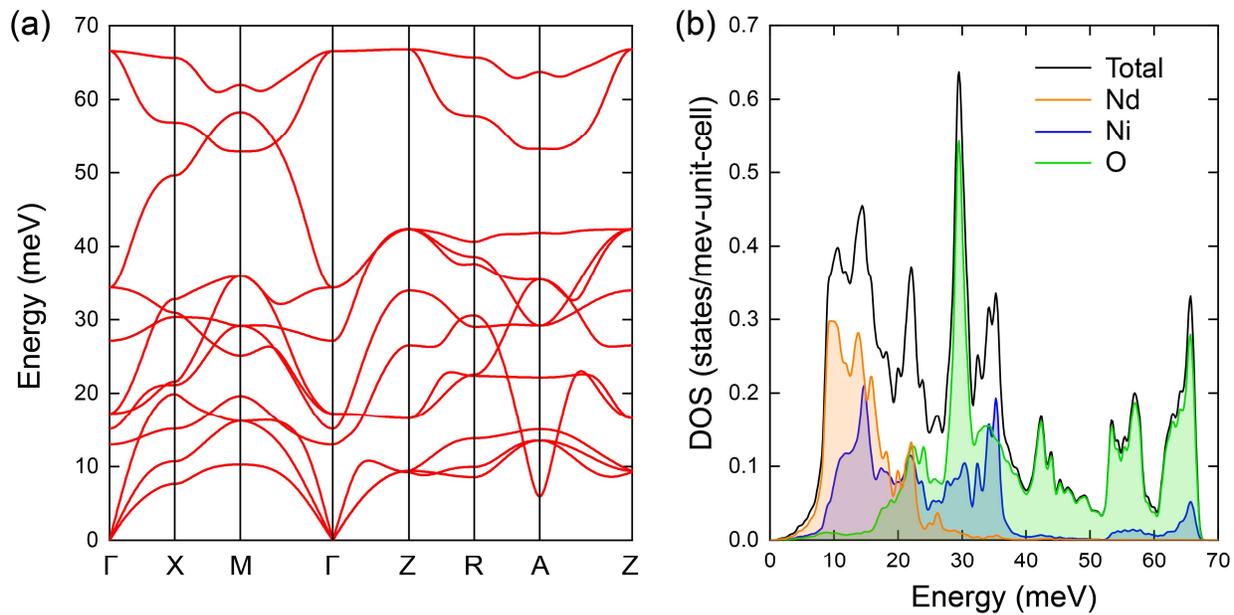

**Fig. S4.** (a) Phonon band structure of $Nd_{0.8}Sr_{0.2}NiO_2$ computed from DFPT. (b) Phonon density of states (DOS), with the total spectrum (black) as well as contributions from projections to Nd (orange), Ni (blue), and O (green) atoms.

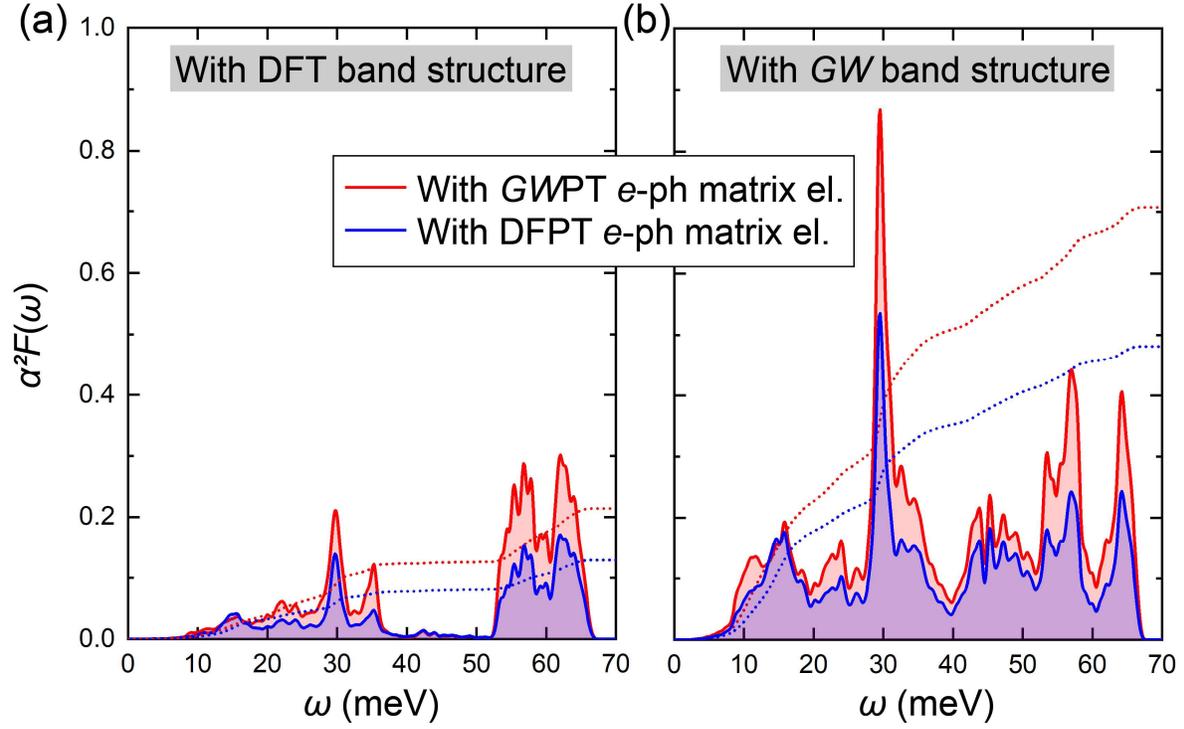

**Fig. S5.** Eliashberg function $\alpha^2 F$ using electronic states from: (a) DFT band structure, and (b) *GW* band structure. With both band structures, we compute results using DFPT and *GW*PT *e*-ph matrix elements. The integrated total *e*-ph coupling constants are $\lambda(\varepsilon^{\text{DFT}}, g^{\text{DFT}}) = 0.13$, $\lambda(\varepsilon^{\text{DFT}}, g^{GW}) = 0.21$, $\lambda(\varepsilon^{GW}, g^{\text{DFT}}) = 0.48$, and $\lambda(\varepsilon^{GW}, g^{GW}) = 0.71$.

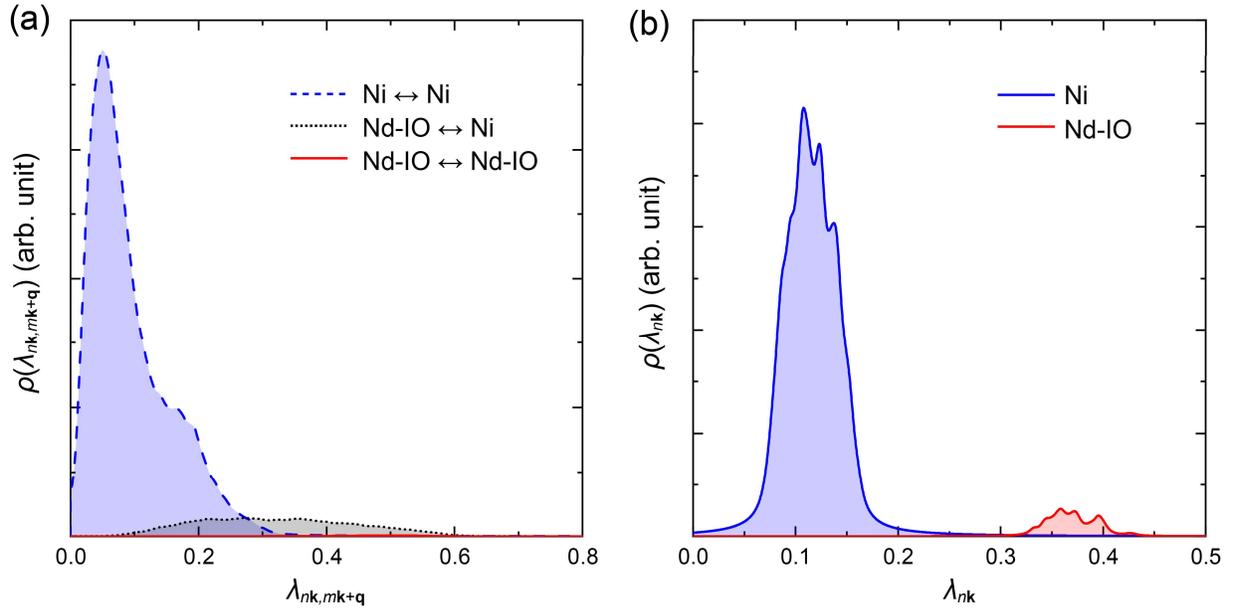

**Fig. S6.** Decomposition of *e*-ph coupling density distribution calculated at the DFT level. (a) Density distribution of the state-pair-resolved *e*-ph coupling $\lambda_{n\mathbf{k},m\mathbf{k+q}}$. (b) Density distribution of the state-resolved *e*-ph coupling $\lambda_{n\mathbf{k}}$. The distributions are for states within a window of $E_F \pm 0.12$ eV.

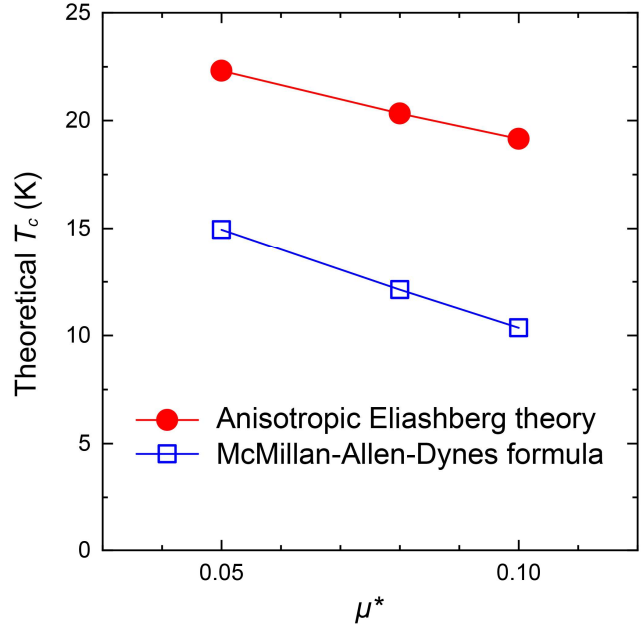

**Fig. S7.** Theoretical $T_c$ of $Nd_{0.8}Sr_{0.2}NiO_2$ at the *GW* level from anisotropic Eliashberg theory and McMillan-Allen-Dynes formula for different values of the effective Coulomb potential $\mu^*$.

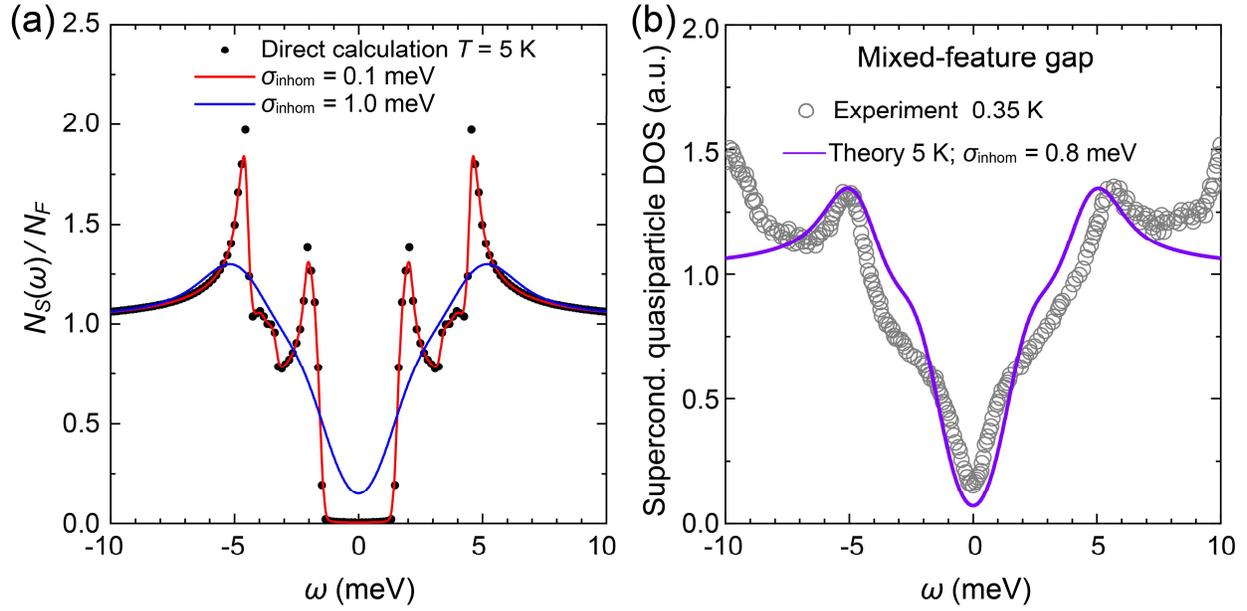

**Fig. S8.** (a) Superconducting quasiparticle DOS (black dots) of $Nd_{0.8}Sr_{0.2}NiO_2$ from solving the anisotropic Eliashberg equations at 5 K at the *GW*-level. Extrinsic smearing (due to crystalline imperfections) is introduced by a broadening parameter $\sigma_{inhom}$ with Gaussian smearing function to the directly calculated DOS data. Small $\sigma_{inhom}$ = 0.1 meV (red curve) well reflects the two-gap structure in the intrinsic DOS whereas large $\sigma_{inhom}$ = 1.0 meV (blue curve) leads to a V-shape DOS. (b) Moderate $\sigma_{inhom}$ = 0.8 meV (purple curve) well reproduces the experimentally observed gap profiles with mixed U-shape and V-shape features [23].

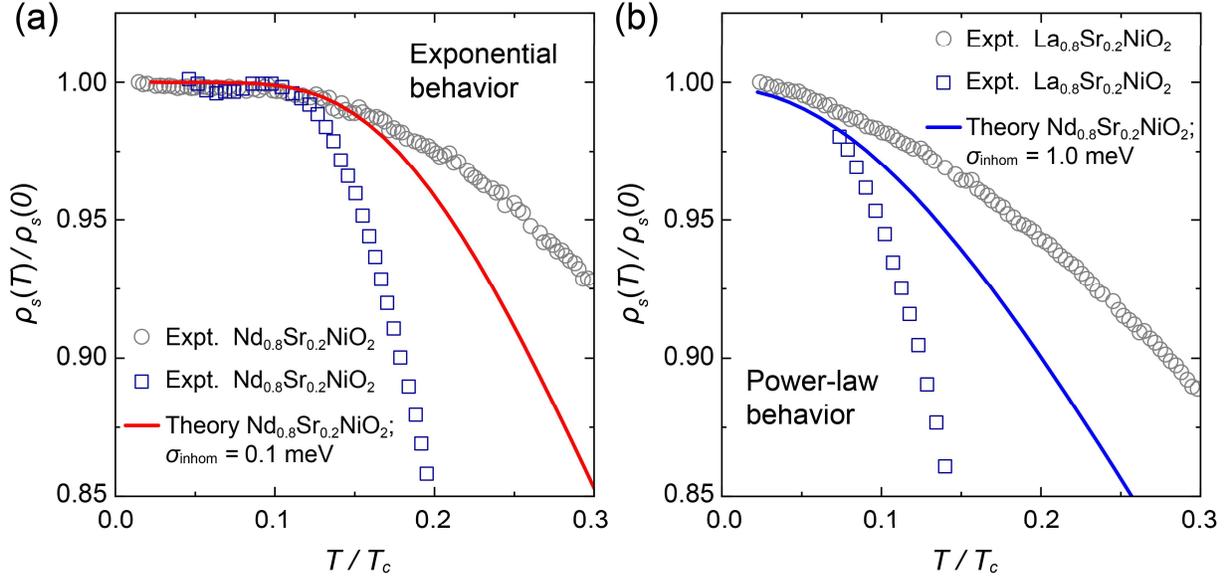

**Fig. S9.** (a) Exponential behavior of the normalized superfluid density $\frac{\rho_s(T)}{\rho_s(0)}$ at low temperature. In the present work, an exponential behavior arises from the computed *s*-wave superconductivity gap with $\sigma_{inhom}$ = 0.1 meV representing the clean limit (with a U-shape superconducting quasiparticle DOS). (b) Power-law behavior of the normalized superfluid density at low temperature. Within the strong scattering regime, $\sigma_{inhom}$ = 1.0 meV (with a V-shape superconducting quasiparticle DOS), a power-law behavior arises from the computed results. Both behaviors (a) and (b) have been observed in infinite-layer nickelate experiments. Circles are the experimental data in Ref. [24] and squares are the experimental data in Ref. [25].

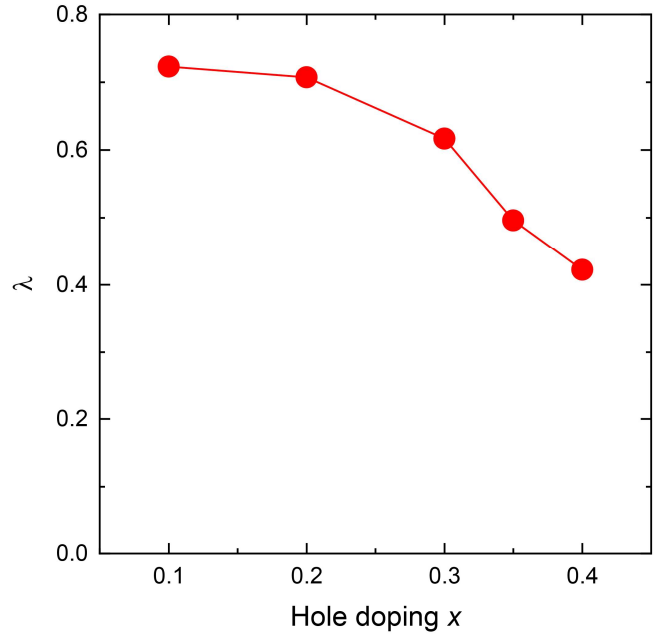

**Fig. S10.** The *GW*-level *e*-ph coupling constant $\lambda$ averaged over the whole Fermi surface as a function of hole doping concentration $x$.